\documentclass[conference]{IEEEtran}
\IEEEoverridecommandlockouts

\usepackage{cite}
\usepackage{amsmath,amssymb,amsfonts}
\usepackage{graphicx}
\usepackage{textcomp}
\usepackage{xcolor}
\usepackage{algorithm}
\usepackage{algpseudocode}
\usepackage{float}
\usepackage{booktabs} 
\usepackage{setspace}

\usepackage{siunitx} 
\usepackage{subfig}
\usepackage[draft=true]{hyperref}
\hypersetup{bookmarks=false}

\graphicspath{ {./figures/} }
\def\BibTeX{{\rm B\kern-.05em{\sc i\kern-.025em b}\kern-.08em
    T\kern-.1667em\lower.7ex\hbox{E}\kern-.125emX}}
\begin{document}

\setstretch{.95}
\title{\Huge{Interference-Aware Queuing Analysis for Distributed Transmission Control in UAV Networks}
}

\author{

\IEEEauthorblockN{Masoud Ghazikor\IEEEauthorrefmark{1},
Keenan Roach\IEEEauthorrefmark{2},
Kenny Cheung\IEEEauthorrefmark{2},
Morteza Hashemi\IEEEauthorrefmark{1}}
        \IEEEauthorblockA{
        \IEEEauthorrefmark{1}Department of Electrical Engineering and Computer Science, University of Kansas \\
        \IEEEauthorrefmark{2}Universities Space Research Association (USRA)
        }
}

\maketitle

\begin{abstract}
In this paper, we investigate the problem of distributed transmission control for unmanned aerial vehicles (UAVs) operating in unlicensed spectrum bands. We develop a rigorous interference-aware queuing analysis framework that \emph{jointly} considers two inter-dependent factors: (i)  limited-size queues with delay-constrained packet arrival, and (ii) in-band interference introduced by other ground/aerial users. We aim to optimize the expected throughput by jointly analyzing these factors. In the queuing analysis, we explore two packet loss probabilities including, buffer overflow model and time threshold model. For interference analysis, we investigate the outage probability and packet losses due to low signal-to-interference-plus-noise ratio (SINR). We introduce two algorithms namely, Interference-Aware Transmission Control (IA-TC), and Interference-Aware Distributed Transmission Control (IA-DTC). These algorithms maximize the expected throughput by adjusting transmission policies to balance the trade-offs between packet drop from queues vs. transmission errors due to low SINRs.  We implement the proposed algorithms and demonstrate that the optimal transmission policy under various scenarios is found.   

\end{abstract}


\section{Introduction}
Unmanned Aerial Vehicles (UAVs) have emerged as a transformative technology for a wide range of applications such as environmental conservation, emergency services, delivery, and more. These instances are crucial to determining how to support the expected increase in UAV usage \cite{Badnava-2021-Spectrum}. For UAV communications, both licensed and unlicensed spectrum can be used.
 Licensed spectrum grants exclusive access to the channel and includes regulatory requirements to fulfill. In contrast, unlicensed spectrum is shared among different communication nodes, and thus it includes light regulations that make
nodes more prone to interference from other users. 
As a result, reliable and robust UAV communication in unlicensed spectrum bands is challenging. To address this issue, developing a distributed transmission policy is essential to ensure a high quality of service, particularly for UAV networks that require delay-sensitive  command-and-control (C2) data. 

Over the past few years, there has been extensive amounts of research on different aspects of UAV networking (see, for example,~\cite{Meng-2022-Beamforming,Badnava-2023-Computing,Sravan-2023-Collaborative}).  
However, there are still research gaps in developing distributed transmission policies 
that jointly take into account \textbf{(i)} the level of interference in the unlicensed spectrum bands, and \textbf{(ii)} the transmission queue state in terms of buffer size and queuing delay. 
For instance, in \cite{Bithas-2023-Generalized},
   Line-of-Sight (LoS) and Non-Line-of-Sight (NLoS) wireless links  are modeled for UAVs and a transmission policy is defined without considering interference. 
 In \cite{Guan-2016-ToTransmit}, distributed transmission policy is developed for ground level terrestrial networks, and not specifically for UAV networks. 
  A new LoS channel model is proposed in \cite{Bithas-2020-Channel} for UAVs, but the interference model is not included in the outage probability calculation. In \cite{Cui-2019-Data}, ground-to-air (G2A) and air-to-ground (A2G) channels are modeled, and system performance is characterized in terms of UAV trajectory, and not the channel parameters. 


\begin{figure}
    \centering
   \includegraphics[width=\linewidth]{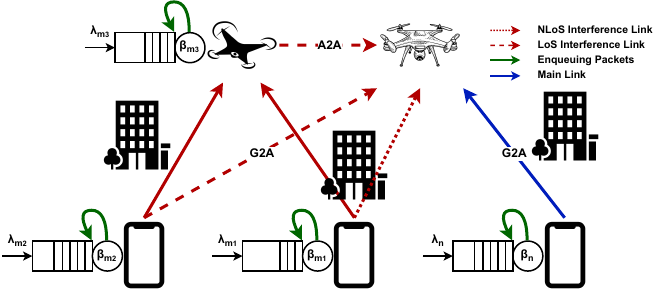}
   \vspace{-.3cm}
    \caption{\small{System model that consists of ground and aerial nodes. }}
    \label{Fig. 1}
    \vspace{-18pt}
\end{figure}

Although there have been research on UAV network transmission policies~\cite{Bithas-2023-Generalized,Bithas-2020-Channel,Cui-2019-Data}, an in-depth investigation of distributed transmission policy that takes into account both packet queues and interference levels in unlicensed frequency bands is still lacking. This paper aims to address this gap by exploring how UAVs and ground-level nodes achieve an optimal policy by adjusting the channel fading threshold in a distributed manner. To this end, we propose a comprehensive analytical framework to characterize two factors that impact packet losses: \textbf{(i) } queue-related analysis wherein we consider a time threshold model to capture the delay sensitivity of data packets, and buffer overflow model to capture the limited size of data queues, and \textbf{(ii)} channel-related analysis by which we focus on the outage probability and analyze the impact of in-band interference on packet transmission errors.

Given these factors, we introduce two transmission algorithms, namely, Interference-Aware Transmission Control (IA-TC), and Interference-Aware Distributed Transmission Control (IA-DTC). In the IA-TC algorithm, we assume there is a single source node and devise a solution to find its optimal channel fading threshold $\beta_n$ and the best expected throughput ($R_n^{best}$), while adjusting the $\beta_m$ of other aerial or ground interferer nodes. 
On the other hand, in the IA-DTC algorithm, we find the optimal channel fading threshold for all nodes, i.e., $\boldsymbol{\beta} \triangleq \left\{\boldsymbol{\beta}_m,\beta_n\right\}$, by assuming that each node can be considered as a source node. 
In this algorithm, each node discovers its optimal $\beta$ through a consensus-based optimization approach.
Through numerical results, we compare the performance of our proposed algorithms with several baselines to demonstrate the efficacy of the our method. 
In summary, the main contributions of this paper are as follows: 
\begin{itemize}
    \item We propose a comprehensive analytical framework for developing throughput-optimal transmission policies for UAV networks. Our system takes into account the effects of queue and in-band interference on packet losses.  
    \item We introduce two transmission policy algorithms for finding the optimal channel fading threshold in the presence of queue and channel impairments. We implement coordinate descent optimization to find the optimal channel fading threshold for the source node and consensus-based distributed optimization to determine the optimal channel fading threshold for each node.
    \item We implement the proposed algorithms and assess their effectiveness under various scenarios in comparison to several alternative baseline policies. 
\end{itemize} 

\noindent The rest of this paper is organized as follows. In Section \ref{system model}, the system model is introduced.
Section \ref{problem formulation} offers an analysis of the time threshold and buffer overflow models, as well as the outage probability by considering the impact of interference. In Section \ref{algorithms}, the IA-TC and IA-DTC transmission policy algorithms are introduced. Section \ref{numerical results} provides the numerical results, followed by the conclusion in Section \ref{conclusion}. 

\section{System Model} \label{system model}
We consider a network that consists of ground and aerial nodes, all operating in unlicensed spectrum bands for G2A, A2G, G2G, and A2A links. The spectrum band is divided into a set of $F$ frequency channels. We further assume that the source node communicates with the \emph{main UAV} (right UAV in Fig. \ref{Fig. 1}), while several interferer ground nodes communicate with the interferer UAV (left UAV in Fig. \ref{Fig. 1}). In general, 
$N$ denotes the set of communication sessions that share the same spectrum band, and $n \in N$ represents the individual session between the source node and the UAV.  

Furthermore, we assume that each node has a limited-size queue, where the packet arrival process ($\lambda_i$) follows a Poisson distribution. Each node either transmits the packet to its destination or keeps the packet in its queue. This transmission decision is determined based on channel conditions and queue states. For instance, if two or more nodes choose the same channel to transmit packets simultaneously, there would be in-band interference and degraded SINR values.

\noindent 
\textbf{Channel Modeling.} Given the described system model, first we calculate the LoS probability ($P_{LoS}(d_i)$) that captures different types of channels \cite{Kim-2019-Impact}:
\begin{align*}
P_{LoS}(d_i) = 
\begin{cases}
    \Bigl(1-e^{(-\frac{z_{i}^2}{2\zeta^2})}\Bigl)^{d_i\sqrt{v\mu}}       & \text{$z_{i} = z_{u}$}\\
    \Bigl(1-\frac{\sqrt{2\pi}\zeta}{d_i^{V}}\left|Q(\frac{z_{i}}{\zeta})-Q(\frac{z_{u}}{\zeta})\right|\Bigl)^{d_i^{H}\sqrt{v\mu}}          & \text{$z_{i} \neq z_{u}$}, 
\end{cases}
\end{align*}
where $\zeta$, $v$, and $\mu$ are environmental parameters and $Q(x)$ is the $Q$-function. Also, $d_i^{H} = \sqrt{(x_{i}-x_{u})^2 + (y_{i}-y_{u})^2}$ and $d_i^{V} = \sqrt{(z_{i}-z_{u})^2}$ are, respectively,  horizontal and vertical distances between the transmitter ($i$) and receiver ($u$). 
Thus, the total distance between node $i$ and a specific receiver node is obtained as:
\begin{align}
d_i = \sqrt{{d_i^H}^2 + {d_i^V}^2} \quad \forall{i} = \left\{n,\boldsymbol{m}\right\}, 
\end{align}
where the indices $n$ and $\boldsymbol{m}$ denote the source node and the set of interferer nodes in a given area, respectively.

Given the transmit power $P_t$, we have $P_r = P_t |h_n^f|^2$ in which $P_r$ is the received power and $h_n^f$ is the channel gain of channel $f \in F$. Furthermore, $h_n^f$ can be expressed as $h_n^f = \Tilde{h}_n^f \hat{h}_n^f$, in which $\Tilde{h}_n^f$ denotes  the channel fading coefficient, and $\hat{h}_n^f$ is the square root of the path loss. By using a single-slope path loss model\cite{Ren-2011-Modelling}, we have: 
\begin{align}
\hat{h}_n^f = \sqrt{c(\frac{d_0}{d_i})^{\alpha(d_i)}} \quad \text{if} \enskip d_i \ge d_0, 
\end{align}
where $c = \frac{\lambda^2}{16\pi^2d_0^2}$ is a constant factor. Furthermore,  $d_0$ and $d_i$ are the reference distance and the distance between the nodes and their intended receiver, respectively. Also, the path loss exponent $\alpha(d_i)$ is defined as \cite{Azari-2018-Ultra,Kim-2018-Outage}:
\begin{align}
\alpha(d_i) = \alpha_{LoS}P_{LoS}(d_i) + \alpha_{NLoS}(1 - P_{LoS}(d_i)), 
\end{align}
in which $\alpha_{LoS}$ and $\alpha_{NLoS}$ are the path loss exponents for LoS and NLoS links, respectively. 

In the framework of a block fading channel model, the variable $\Tilde{h}_{n}^f$ follows either the Rician (Rice) or Rayleigh (Ray) distributions depending on whether it corresponds to LoS or NLoS channels, respectively. Let us initially focus on the Rician channel, in which the probability density function (PDF) is given by:
\begin{align}
Pb(\Tilde{h}_{n}^f = x) = xe^{-\frac{x^2+b^2}{2}}I_0(xb), 
\end{align}
where $b = \sqrt{2K(d_i)}$ is defined according to the Rician factor $K(d_i) = K_{NLoS}e^{\ln(\frac{K_{LoS}}{{K_{NLoS}}})P_{LoS}(d_i)^2}$ in which $K_{LoS}$ and $K_{NLoS}$ are determined when $P_{LoS}(d_i)$ is equal to one and zero, respectively \cite{Kim-2019-Impact}. Also, $I_0$ represents the modified Bessel function of the first kind with order zero.

\noindent 
\textbf{Principle of the Transmission Policy. } The source node transmits its packet to the UAV over the best frequency channel $f^* = {\arg\max}_{f \in F} \ \ \Tilde{h}_n^f \hat{h}_n^f$ if the channel fading coefficient is larger than a channel fading threshold $\beta_n$, i.e., $\Tilde{h}_n^{f^*} \ge \beta_n$; otherwise,  the source node would enqueue the packet \cite{Guan-2016-ToTransmit}. Let $\beta_n>0$ be the channel fading threshold.  Based on $\beta_n$, the cumulative distribution function (CDF) of the Rician distribution is given by:
\begin{align}
\begin{aligned}
Pb(\Tilde{h}_{n}^f < \beta_n^{Rice}) = \int_{0}^{\beta_n^{Rice}} xe^{-\frac{x^2+b^2}{2}}I_0(xb) \,dx
\\
= 1-Q_1(b,\beta_n^{Rice}), 
\end{aligned}
\end{align}
where $Q_1$ represents the first-order Marcum $Q$-function. Therefore, assuming that $|F|$ represents the cardinality of the set $F$, the transmission probability of the source node during a time slot over the Rician channel can be expressed as:
\begin{align} \label{murice}
\begin{aligned}
\mu_n(\beta_n^{Rice}) = 1-Pb(\Tilde{h}_n^{f^*}<\beta_n^{Rice}) 
\\
= 1-(1-Q_1(b,\beta_n^{Rice}) )^{|F|}. 
\end{aligned}
\end{align}
The same approach can be applied to the Rayleigh channel, where the PDF of the Rayleigh distribution is given by:
\begin{align}
Pb(\Tilde{h}_{n}^f = x) = \frac{2x}{\Omega}e^{-\frac{x^2}{\Omega}}, 
\end{align}
where $\Omega$ is the Rayleigh fading factor. Then, the CDF of the Rayleigh distribution is formulated as follows:
\begin{align*}
\begin{aligned}
Pb(\Tilde{h}_{n}^f < \beta_n^{Ray}) = \int_{0}^{\beta_n^{Ray}} \frac{2x}{\Omega}e^{-\frac{x^2}{\Omega}} \,dx= 1-e^{-\frac{(\beta_n^{Ray})^2}{\Omega}}.
\end{aligned}
\end{align*}
Finally, the transmission probability of a packet from source node in a time slot over the Rayleigh channel is defined as:
\begin{align} \label{muray}
\mu_n(\beta_n^{Ray}) = 1-(1-e^{-\frac{(\beta_n^{Ray})^2}{\Omega}})^{|F|}. 
\end{align}
Given the presented system model, our goal is to find the optimal values for the channel fading threshold $\beta_n$ such that the network throughput is maximized.  

\section{Problem Formulation} \label{problem formulation}
In this section, we characterize throughput performance in terms of its constituent queuing and interference components. 

\subsection{Queuing Characterization} 
We focus on two queue management mechanisms designed to regulate the number of packets in the queue.

\noindent 
\textbf{Time Threshold Model.} UAVs may communicate delay-sensitive data such as command-and-control messages. In this case, it is critical to ensure that data packets are delivered to their intended destination in a specified timeout value.
Assume that $T_n$ denotes the waiting time in the queue for node $n$.  
When the source node is unable to transmit packets due to poor channel conditions (e.g., low SINR), any packet with a waiting time $T_n$ greater than the time threshold $T_n^{th}$ is discarded.

By using $\mu_n(\beta_n)$ derived from Eq.~\eqref{murice} and Eq. \eqref{muray} for Rician and Rayleigh channels in a M/M/1 queue scenario, the probability of packet loss due to exceeding the time threshold can be expressed as~\cite{Guan-2016-ToTransmit}:
\begin{align}
P_n^{dly} (\beta_n) \triangleq Pb(T_n > T_n^{th}) = e^{-(\frac{\mu_n(\beta_n)} {T_{slt}} - \lambda_n) T_n^{th}}, 
\end{align}
where $T_{slt}$ and $\lambda_n$ denote the time slot duration and average incoming packet rate, respectively.
In order to determine the upper bound for $\beta_n$, an important parameter in our IA-TC and IA-DTC algorithms, it is known that $P_n^{dly} (\beta_n) \le 1$. Consequently, the upper bounds for $\beta_n$ in the case of Rician and Rayleigh channels can be obtained as:
\begin{align}
\begin{cases}
    Q_1(b,\beta_n^{Rice}) \ge 1-(1-\lambda_n T_{slt})^{\frac{1}{|F|}},       & \text{Rician;}\\
    \beta_n ^ {Ray} \le \sqrt{-\Omega ln[1-(1-T_{slt} \lambda_n) ^ \frac{1}{|F|}]}         & \text{Rayleigh.}
\end{cases}
\end{align}
The authors in \cite{Guan-2016-ToTransmit} have provided the derivations under Rayleigh channel conditions.

\noindent 
\textbf{Buffer Overflow Model.} In addition to time-threshold model that captures time-sensitivity of data traffics, assume that queues have limited buffer sizes as well.  Therefore, there are chances that new packet arrivals are inadmissible due to buffer overflow, and thus they are dropped. 
By applying the principles of queuing theory, the probability of exceeding the buffer capacity in a certain state $i$ can be defined \cite{Gross-2008-Fundamentals, Ghazikor-2023-Exploring}:
\begin{align*}
\overline{P_{i,i+1}} = P[X_1+...+X_{i+1} > B_n | X_1+...+X_i \le B_n] \nonumber
\\
= \frac{\int_0^{B_n}P[X_{i+1} > B_n - x]f_{X_1+...+X_i}(x)dx}{P[X_1+...+X_i \le B_n]}, 
\end{align*}
in which $B_n$ and $X$ denote the buffer capacity and the packet length, respectively. For the sake of analysis, we assume that the packet length follows an exponential random variable with a parameter $\eta_n$. Also, $f_{X_1+...+X_i}(x)$ represents the PDF of an i-Erlang distribution, respectively. Hence, the complement of $\overline{P_{i,i+1}}$ without occurring buffer overflow is defined as:
\begin{align}
P_{i,i+1} = 1 - \overline{P_{i,i+1}} = \frac{1 - \sum_{j=0}^i \frac{(B_n\eta_n)^j}{j!}e^{-B_n\eta_n}}{1 - \sum_{j=0}^{i-1} \frac{(B_n\eta_n)^j}{j!}e^{-B_n\eta_n}}. 
\end{align}
According to the Markov chain, the local balance equation is $\pi_{i+1} = \rho_n(\beta_n) P_{i,i+1}\pi_{i}$, where $\rho_n(\beta_n) = \frac{\lambda_nT_{slt}}{\mu_n(\beta_n)}$ is the offered load. Then, $\pi_i$ can be derived as:
\begin{align}
\begin{aligned}
\pi_i = \rho_n^i(\beta_n) \bigg(\prod_{j=0}^{i-1}P_{j,j+1}\bigg)\pi_0 = 
\\
\rho_n^i(\beta_n) \bigg(1-\sum_{j=0}^{i-1}\frac{(B_n\eta_n)^j}{j!}e^{-B_n\eta_n}\bigg)\pi_0. 
\end{aligned}
\end{align}
The probability of buffer overflow can be approximated by:
\begin{align}
P_n^{ov}(\beta_n) \approx \sum_{i=0}^\infty \overline{P_{i,i+1}}\pi_i = \frac{(1-\rho_n(\beta_n))e^{-B_n\eta_n(1-\rho_n(\beta_n))}}{1-\rho_n(\beta_n) e^{-B_n\eta_n(1-\rho_n(\beta_n))}}. 
\end{align}
Next, we characterize the impacts of interference as a function of the transmission policy parameter $\beta_n$. 

\subsection{Interference Characterization} \label{interference analysis}
In this part, we investigate the impact of interference on the UAV using the signal-to-interference-plus-noise ratio (SINR). If the SINR falls below the SINR threshold $\gamma_{th}$, a transmission error occurs. 
Let $I_n^f(\boldsymbol{\beta}_{-n})$ be the impact of the interferer nodes on the UAV~\cite{Guan-2016-ToTransmit}:
\begin{align} \label{interference}
I_n^f(\boldsymbol{\beta}_{-n}) = \sum_{m\in N \backslash n} P_m(\hat{h}_{mn}^f\Tilde{h}_{mn}^f)^2\alpha_m^f(\beta_m), 
\end{align}
where the channel fading threshold of the interferer nodes is defined as $\boldsymbol{\beta}_{-n} \triangleq (\boldsymbol{\beta}_m)_{m \in N \backslash n}$. Also, $\alpha_m^f(\beta_m)$ equals one if interferer node $m$ transmits using channel $f$, and zero otherwise. Thus, the outage probability  is defined as:
\begin{align}
P_n^{out}(\boldsymbol{\beta}) \triangleq Pb(\gamma_n < \gamma_{th}) = Pb\Biggl(\frac{P_n(\hat{h}_n^f)^2(\Tilde{h}_n^f)^2}{\sigma^2+I_n^f(\boldsymbol{\beta}_{-n})} < \gamma_{th} \Biggl). \nonumber
\end{align}
Here, $P_n$ is the transmission power, $\sigma^2  = kTW$ denotes the thermal noise power where $k$, $T$, and $W$ are the Boltzmann’s constant, temperature, and bandwidth, respectively. 

By adopting a classical stochastic geometry approach to model $I_n^f(\boldsymbol{\beta}_{-n})$ using Gamma distribution, the final expression for the outage probability is given by~\cite{Guan-2016-ToTransmit}:
\begin{align}
\begin{aligned}
P_n^{out}(\boldsymbol{\beta}) = \int_{\beta_n}^\infty Pb(\Tilde{h}_{n}^f = x) v_n(\frac{P_n (\hat{h}_n^f)^2}{\gamma_{th}} x^2 - \sigma^2,\boldsymbol{\beta}_{-n})dx, 
\end{aligned}
\end{align}
where $Pb(\Tilde{h}_{n}^f = x)$ is determined according to the channel's type (Rician or Rayleigh) and $v_n(x,\boldsymbol{\beta}_{-n})$ is the complementary cumulative distribution function (CCDF) of $I_n^f(\boldsymbol{\beta}_{-n})$. From \cite{Guan-2016-ToTransmit}, we have: 
\begin{align*}
\begin{aligned}
v_n(x,\boldsymbol{\beta}_{-n}) = Pb(I_n^f(\boldsymbol{\beta}_{-n})>x)
= 1-\frac{\varphi(k_n(\boldsymbol{\beta}_{-n}),\frac{x}{\theta_n(\boldsymbol{\beta}_{-n})})}{\Gamma(k_n(\boldsymbol{\beta}_{-n}))}, 
\end{aligned}
\end{align*}
where $\varphi(k_n(\boldsymbol{\beta}_{-n}),\frac{x}{\theta_n(\boldsymbol{\beta}_{-n})}) = \int_0^{\frac{x}{\theta_n(\boldsymbol{\beta}_{-n})}}s^{k_n(\boldsymbol{\beta}_{-n})-1}e^{-s}ds$ is the lower incomplete gamma function and $\Gamma(k_n(\boldsymbol{\beta}_{-n})) = \int_0^{\infty}x^{k_n(\boldsymbol{\beta}_{-n})-1}e^{-x}dx$ is the Gamma function.

\subsection{Throughput Characterization}
Given the presented queuing and interference analysis, now we consider all three packet loss probabilities, namely: {(i)} packet loss due to time threshold $P_n^{dly}(\beta_n)$, {(ii)} packet loss due to buffer overflow $P_n^{ov}(\beta_n)$, and {(iii)} packet loss due to outage and low SINR $P_n^{out}(\boldsymbol{\beta})$. Thus, the probability of overall loss $P_n^{loss}(\boldsymbol{\beta})$ is determined as:
\begin{align}
\begin{aligned}
P_n^{loss}(\boldsymbol{\beta}) = P_n^{ov}(\beta_n) + [1-P_n^{ov}(\beta_n)]P_n^{dly}(\beta_n) + 
\\
[1-P_n^{ov}(\beta_n)][1-P_n^{dly}(\beta_n)]P_n^{out}(\boldsymbol{\beta}). 
\end{aligned}
\end{align}
Since the products of $P_n^{dly}(\beta_n), P_n^{ov}(\beta_n),$ and $P_n^{out}(\boldsymbol{\beta})$ are negligible, we consider only the ``first order'' terms, and thus the expected throughput can be approximated as~\cite{Guan-2016-ToTransmit}:
\begin{align}
\begin{aligned}
R_n(\boldsymbol{\beta}) = \lambda_n [1-P_n^{loss}(\boldsymbol{\beta})] \approx 
\\
\lambda_n [1-P_n^{dly}(\beta_n)-P_n^{ov}(\beta_n)-P_n^{out}(\boldsymbol{\beta})].
\end{aligned}
\end{align}
Next, we present two transmission policies to maximize the expected throughput performance by finding the optimal channel fading threshold $\beta$. 

\section{Proposed Transmission Policy Algorithms} \label{algorithms}

\textbf{Interference-Aware Transmission Control.}
As mentioned, our goal is to develop a transmission policy that achieves the maximum expected throughput for the source node. To this end, we consider aim to
$
\mathop{\mathrm{max}}_{\boldsymbol{\beta}} R_n(\boldsymbol{\beta})$, subject to the upper bound constraints on $\beta_m^{Rice}$ and $\beta_m^{Ray}$, which should be smaller than  $\beta^{Rice}_{max}$ and $\beta^{Ray}_{max}$, respectively. To solve this problem, we use a coordinate descent algorithm to maximize a function ($R_n(\boldsymbol{\beta})$) by adjusting different variables (i.e., components of the vector $\boldsymbol{\beta}$) across each coordinate separately. 

At each step, the algorithm focuses on one variable and updates it, while keeping the other variables constant. In particular, Algorithm \ref{alg1} optimizes the expected throughput ($R_n$) by adjusting the channel fading thresholds $\boldsymbol{\beta} \triangleq \left\{\boldsymbol{\beta}_m,\beta_n\right\}$ for the interferer ($\boldsymbol{\beta}_m \triangleq \left\{\beta_m^{Rice},\beta_m^{Ray}\right\} $) and source ($\beta_n$) nodes. Consider the same $\beta_m^{Rice}$, $\beta_m^{Ray}$ for all interferers and $\beta_n$ as three coordinate axes. Initially (lines 5-7), we set $\beta_n$ and $\beta_m^{Ray}$ as fixed parameters and change $\beta_m^{Rice}$ by the step of $stp_m$ to determine the best expected throughput ($R_n^{best}$) in the specified coordinate. Then (lines 8-11), we repeat the same procedure while interchanging $\beta_m^{Rice}$, $\beta_m^{Ray}$, and $\beta_n$.

To control $\boldsymbol{\beta}_m$, we set its maximum value as the smallest upper bound for two types of channels among all interferer nodes, denoted as $\beta^{Rice}_{max}$ and $\beta^{Ray}_{max}$. However, there is no need to set a maximum value for $\beta_n$ as it dynamically adjusts itself through iteration. Also, stopping criteria are based on the maximum number of iterations ($maxiter$) and the condition that the difference between $R_n^{best}$ and the previous one ($R_n^{prev}$) is lower than $\epsilon$.
In the Coordinate Search (CS) function (lines 6, 9, and 11), we explore the coordinate by $\boldsymbol{stp}_i \triangleq \left\{stp_m, stp_n\right\} $ to determine $R_n^{best}$ and $\boldsymbol{\beta}$ in each coordinate axis.

\noindent 
\emph{\underline{Remarks.}} IA-TC algorithm constantly tries to increase $\beta_m^{Rice}$ and $\beta_m^{Ray}$ until it reaches the maximum value since the interferer nodes send fewer packets and the level of interference on the main link decreases. Furthermore, as $\beta_n$ increases, the source node enqueues more packets. Thus, while packet loss in the queue rises, packet loss due to transmission error decreases. Therefore, as the number of iterations of the IA-TC algorithm increases, the values of $\beta_m^{Rice}$ and $\beta_m^{Ray}$ increases, and the value of $\beta_n$ decreases. Therefore, the main source node would have more transmission opportunities, while the transmission attempts by the interferer nodes decreases. Our numerical results in Section~\ref{numerical results} confirm this result. 


\begin{algorithm}[t]
\caption{\small{Interference-Aware Transmission Control (IA-TC)}}\label{alg1}
\begin{algorithmic}[1]
    \Function{IA-TC}{$\boldsymbol{\beta}^{ini}$, $\beta_{max}^{Rice}$, $\beta_{max}^{Ray}$, $\boldsymbol{stp}_i$, $maxiter$}
      \State $\boldsymbol{\beta} \gets \boldsymbol{\beta}^{ini}$, $R_n^{best} \gets R_n(\boldsymbol{\beta})$
      \For{$iter$ \textbf{in} \textbf{range} $maxiter$}
        \State $R_n^{prev} \gets R_n^{best}$
        \If{$\exists \, \text{Rice} \in \boldsymbol{m}$ \textbf{and} $\beta_m^{Rice} + stp_m < \beta^{Rice}_{max}$}
            \State $\boldsymbol{\beta}, R_n^{best} = CS(\boldsymbol{\beta}_m, stp_m, R_n^{best}, \beta_n)$
        \EndIf
        \If{$\exists \, \text{Ray} \in \boldsymbol{m}$ \textbf{and}  $\beta_m^{Ray} + stp_m < \beta^{Ray}_{max}$}
            \State $\boldsymbol{\beta}, R_n^{best} = CS(\boldsymbol{\beta}_m, stp_m, R_n^{best}, \beta_n)$
        \EndIf
        \State $\boldsymbol{\beta}, R_n^{best} = CS(\beta_n, stp_n, R_n^{best}, \boldsymbol{\beta}_m)$
        \If{$|R_n^{prev} - R_n^{best}| < \epsilon$}
            \State \textbf{break}
        \EndIf
      \EndFor
      \State \textbf{return} $\boldsymbol{\beta}, R_n^{best}$
    \EndFunction
\end{algorithmic}
\end{algorithm}

\textbf{Interference-Aware Distributed Transmission Control.}
Here, our goal is to develop a distributed transmission policy that achieves the maximum expected throughput across all links, while assuming that each link can be a main link. When compared to IA-TC, increasing the channel fading threshold for interferer nodes is no longer optimal because they could also serve as the main link. In this case, distributed nodes should coordinate with each other to converge to the optimal transmission policy that is desirable for all nodes, rather than just one, as in IA-TC.


We use \emph{consensus-based distributed optimization} to solve this problem in which multiple nodes collaborate to reach a consensus on $\boldsymbol{\beta}$. Each node has its own local information and objective function ($R_n(\boldsymbol{\beta})$), and it communicates iteratively with its neighbors to find optimal $\boldsymbol{\beta}$ and maximize $R_n(\boldsymbol{\beta})$~\cite{Berahas-2019-Balancing}. 

In Algorithm \ref{alg3}, our goal is to determine the optimal set of channel fading thresholds ($\boldsymbol{\beta^{\star}}$) for nodes. Initially, we start by setting the channel fading threshold of interferer nodes to the maximum value, allowing each node to selfishly identify its best channel fading threshold ($\beta_n^{best}$) based on the results obtained from the IA-TC algorithm. During each iteration, if the difference between the updated channel fading threshold set and the previous one is greater than $\epsilon$, nodes exchange information regarding their channel fading thresholds with each other to determine the optimal channel fading threshold.

In this algorithm, the set $\hat{\boldsymbol{m}}$ includes all nodes (source or interferer), also, $n$ and set $\boldsymbol{m}$ specifies the source node and interferer nodes, respectively. We introduce the $\boldsymbol{\beta}^{can}$ list to collect the updated channel fading thresholds as different nodes are selected as the source node. In lines 2-7, the primary $\boldsymbol{\beta} \triangleq (\boldsymbol{\beta}_m, \beta_n)_{m,n \in N}$ is set to the maximum value of the channel fading threshold ($\boldsymbol{\beta}^{max}$) according to the upper bound of it. Then, selfish values of the channel fading threshold are stored in $\boldsymbol{\beta}^{can}$. In lines 8-20, nodes find their best channel fading thresholds by $\boldsymbol{\beta}$ and then update $\boldsymbol{\beta}^{can}$. In each iteration, $R_n(\boldsymbol{\beta})$ for all nodes is stored in $\boldsymbol{R}$. The Local Coordinate Search (LCS) function (lines 5 and 13) determines $\beta_n^{best}$ for each node while having access to $\boldsymbol{\beta}_{m}$. This function explores a coordinate by $stp$ until it finds $\beta_n^{best}$ associated with $R_n^{best}$.

\begin{algorithm}[t]
\caption{\small{Interference-Aware Distributed Transmission Control (IA-DTC)}} \label{alg3}
\begin{algorithmic}[1]
\Function{IA-DTC}{$\boldsymbol{\beta}^{max}$, $\hat{\boldsymbol{m}}$, $stp$, $maxiter$}
    \State $\boldsymbol{m}^{prev} \gets \hat{\boldsymbol{m}}$, $ \boldsymbol{\beta} \gets \boldsymbol{\beta}^{max}$
    \For{$n$ \textbf{in} \textbf{range} $\hat{\boldsymbol{m}}$}
        \State $\beta_n \gets \boldsymbol{\beta}[n]$, $\boldsymbol{m} \gets \hat{\boldsymbol{m}} - \{n\}$
        \State $\beta_n^{best} = LCS(maxiter, stp, \beta_n, \boldsymbol{\beta}_{m})$
        \State $\boldsymbol{\beta}^{can}[n] \gets \beta_n^{best}$, $\hat{\boldsymbol{m}} \gets \boldsymbol{m}^{prev}$
    \EndFor
    \For{$iter$ \textbf{in} \textbf{range} $maxiter$}
        \State $\boldsymbol{\beta} \gets \boldsymbol{\beta}^{can}$
        \For{$n$ \textbf{in} \textbf{range} $\hat{\boldsymbol{m}}$}
            \State $\beta_n \gets \boldsymbol{\beta}[n]$, $\boldsymbol{m} \gets \hat{\boldsymbol{m}} - \{n\}$
            \State $\boldsymbol{R}[iter][n] \gets R_n(\beta_n, \boldsymbol{\beta}_{m})$
            \State $\beta_n^{best} = LCS(maxiter, stp, \beta_n, \boldsymbol{\beta}_{m})$
            \State $\boldsymbol{\beta}^{can}[n] \gets \beta_n^{best}$, $\hat{\boldsymbol{m}} \gets \boldsymbol{m}^{prev}$
        \EndFor
            \If{$|\boldsymbol{\beta} - \boldsymbol{\beta}^{can}| < \epsilon$}
            \State $\boldsymbol{\beta}^{\star} \gets \boldsymbol{\beta}^{can}$
            \State \textbf{break}
        \EndIf
    \EndFor
    \State \textbf{return} $\boldsymbol{\beta}^{\star}$, $\boldsymbol{R}$
\EndFunction
\end{algorithmic}
\end{algorithm}

\section{Numerical Results} \label{numerical results}
In our setup, 10 nodes (1 main UAV, 1 interferer UAV, 8 ground nodes) are placed according to the Poisson distribution in the area, and the main link is established between the source node (node $8$) and the main UAV at 40 m height (node 10), while other ground nodes (nodes 1 to 7) serve as interferer nodes communicating with the interferer UAV (node 9). In our simulations, we assume that the ground nodes experience the same type of channel conditions (Rician or Rayleigh) to the main and interferer UAVs.
Key simulation parameters are summarized in Table \ref{parameters}.

\begin{table}[t]
    \centering
    \caption{Key Simulation Parameters}
    \resizebox{\columnwidth}{!}{%
    \label{tab:sim_parameters}
    \begin{tabular}{lc}
        \toprule
        Parameter & Value \\
        \midrule
        Communication Area & $100 \times 100 \, \text{m}^2$ \\
        $P_{LoS}$ Model & $\zeta = 20$, $v = 3 \times 10^{-4}$, $\mu = 0.5$\\
        Path Loss Model & $\alpha_{LoS}=2$, $\alpha_{NLoS}=3.5$, $d_0=10$ m \\
        Channel Model & $|F|=14$, $\Omega=2$\\
        Rician Factor Model & $K_{LoS}=15$, $K_{NLoS}=1$ \\
        Queuing Model (1) & $T^{th}_n=80$ ms, $T_{slt}=5$ ms\\
        Queuing Model (2) & $\lambda_n=80$, $B_n \eta_n=100$\\
        Interference Model & $P_i=0.5$ W, $\gamma_{th}=10$ \\
        Noise Model & $f = 2.4$ GHz, $T = 290$ K \\
        IA-TC Algorithm (1) & $\boldsymbol{\beta}^{ini}=[3, 2, 4]$, $\boldsymbol{stp}_i=[0.05, 0.02]$ \\
        IA-TC Algorithm (2) & $\beta_{max}^{Rice}=4.08$, $\beta_{max}^{Ray}=2.57$ \\
        IA-DTC Algorithm & $stp=0.05$, $maxiter=100$ \\
        \bottomrule
    \end{tabular}
    }
    \label{parameters}
\end{table}



\begin{figure*}[t] 
    \centering
    \begin{minipage}{.33\textwidth} 
        \includegraphics[width=6cm]{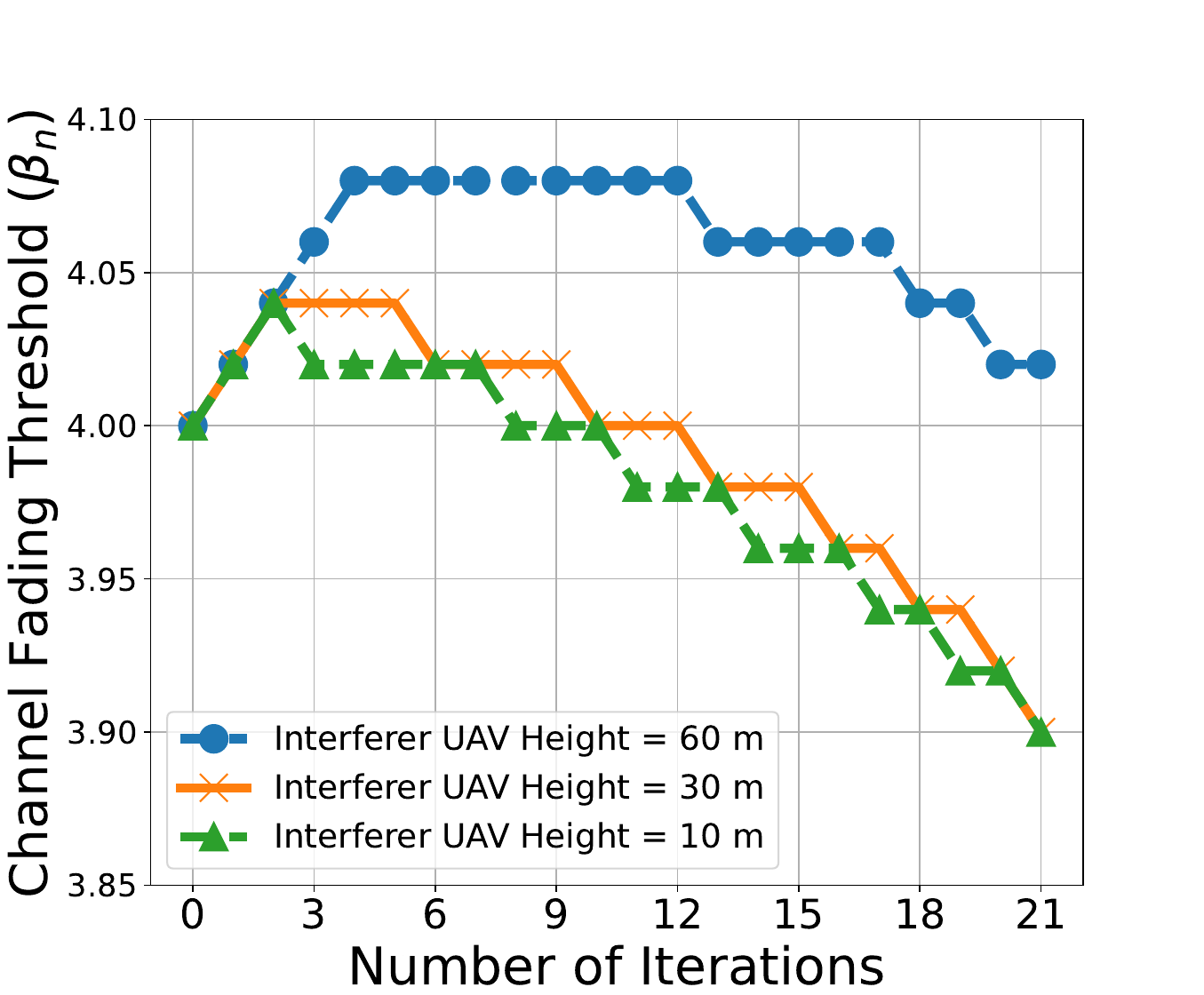}
        \caption{$\beta_n$ vs. \# of Iterations by IA-TC}
        \label{fig2}
    \end{minipage}%
    \begin{minipage}{.33\textwidth} 
        \includegraphics[width=6cm]{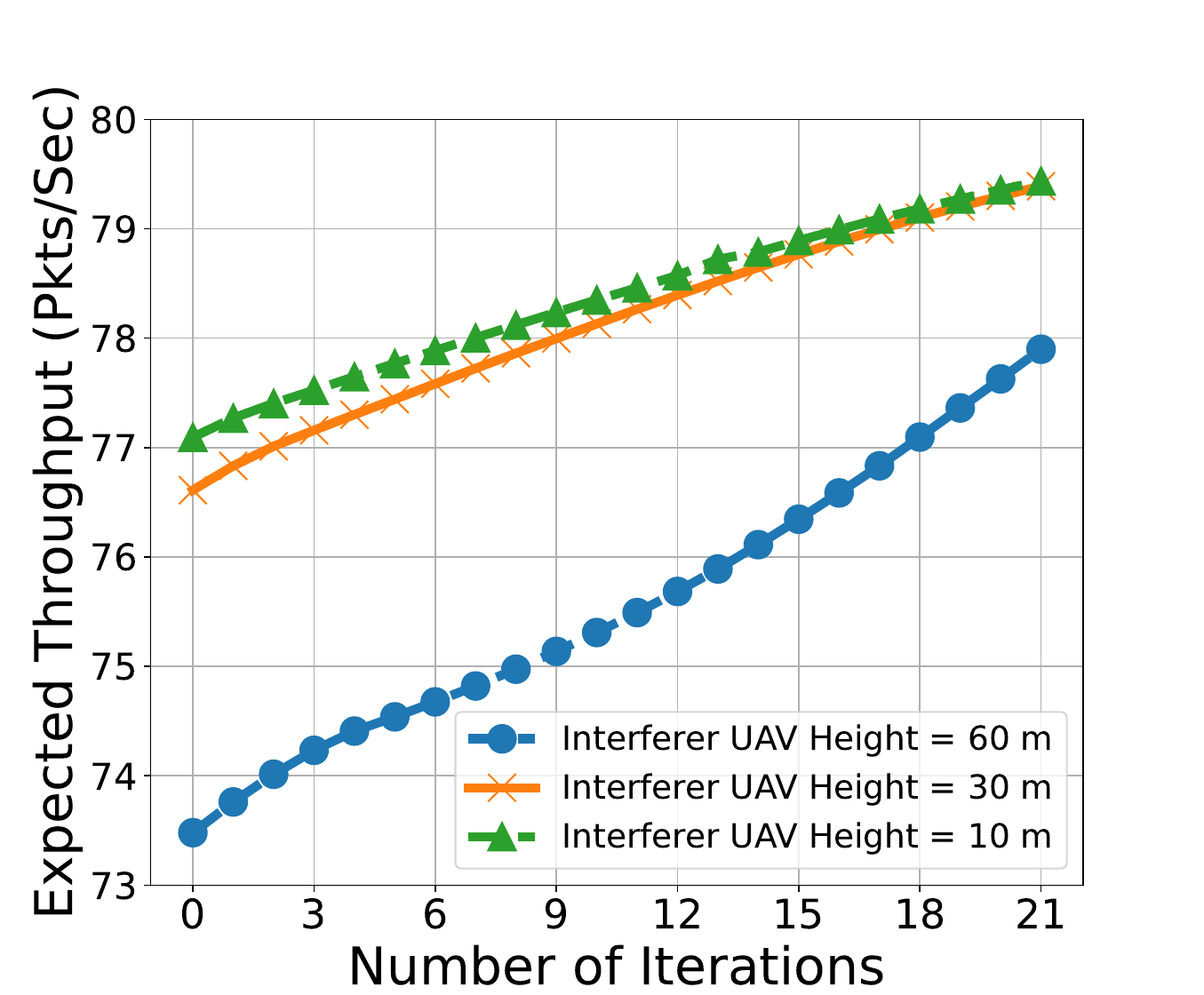}
        \caption{$R_n$ vs. \# of Iterations by IA-TC}
        \label{fig3}
    \end{minipage}
    \begin{minipage}{.33\textwidth} 
        \includegraphics[width=6cm]{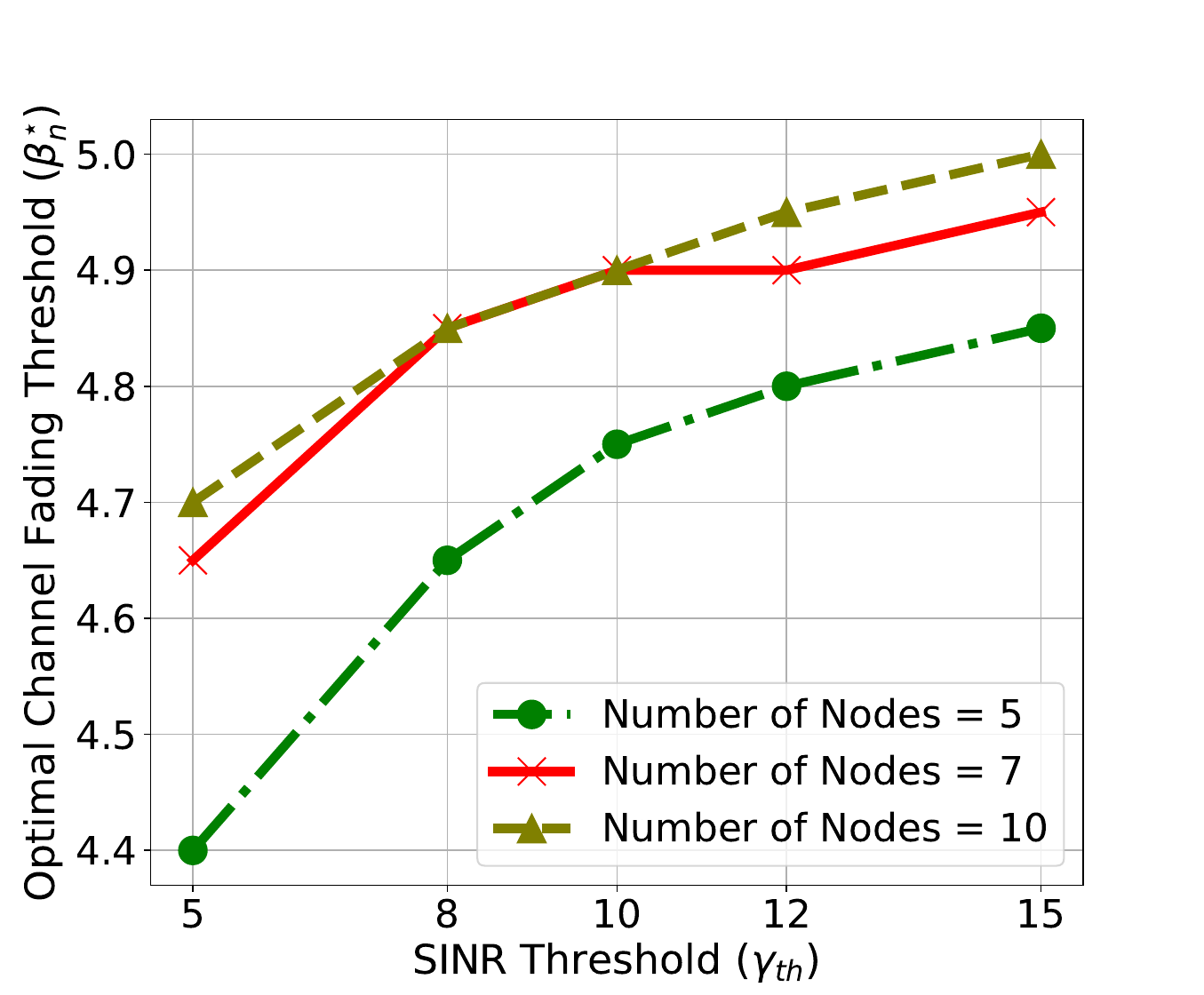}
        \caption{$\beta_n^{\star}$ vs. $\gamma_{th}$ by IA-DTC}
        \label{fig4}
    \end{minipage}
    \vspace{-.4cm}
    \label{fig:example}
    \label{IA-DTC1}
\end{figure*}

\begin{figure*}[t] 
    \centering
    \begin{minipage}{.33\textwidth} 
        \includegraphics[width=5.5cm]{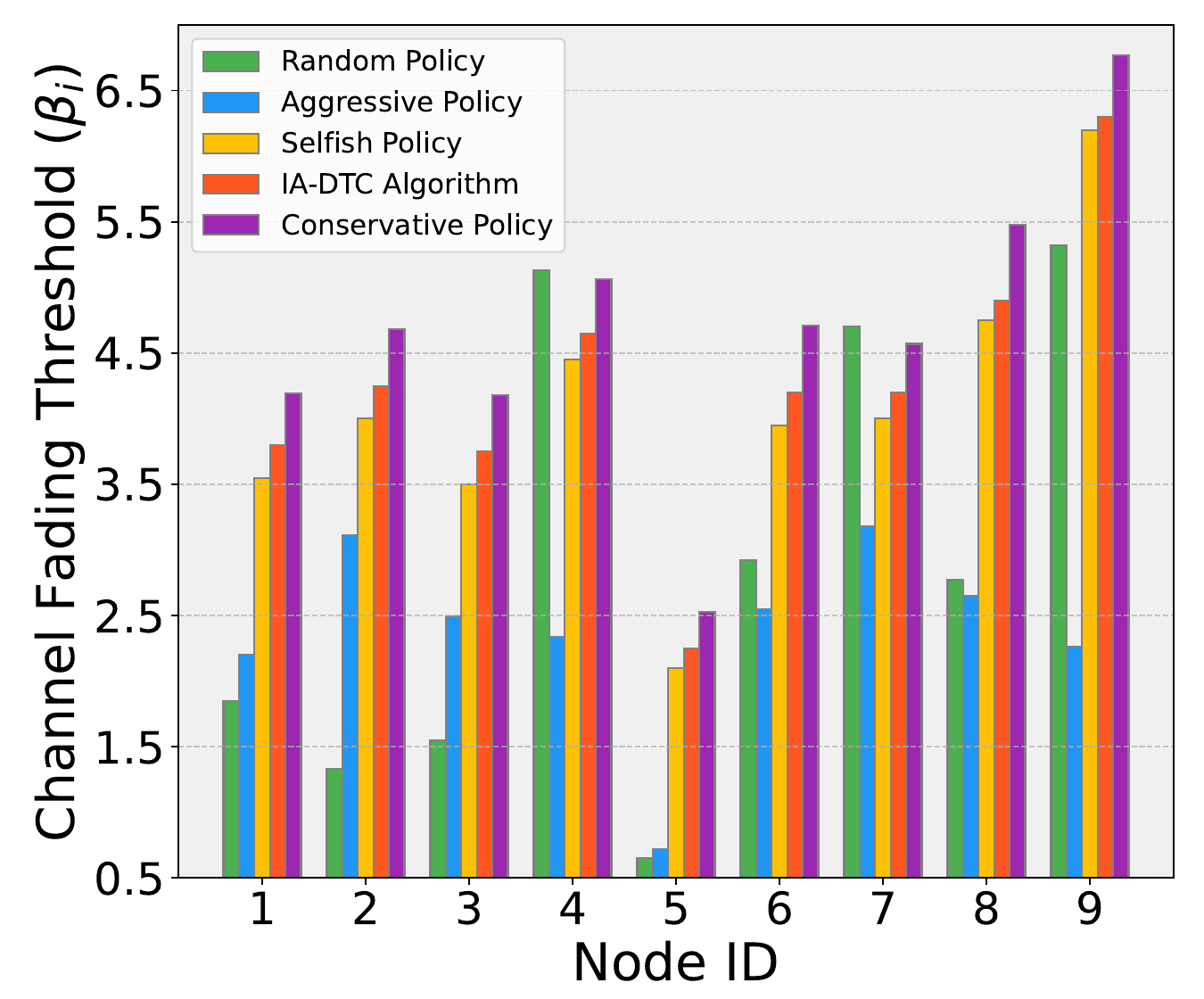}
        \caption{$\beta_i$ vs. Node ID by IA-DTC}
        \label{fig5}
    \end{minipage}%
    \begin{minipage}{.33\textwidth} 
        \includegraphics[width=5.5cm]{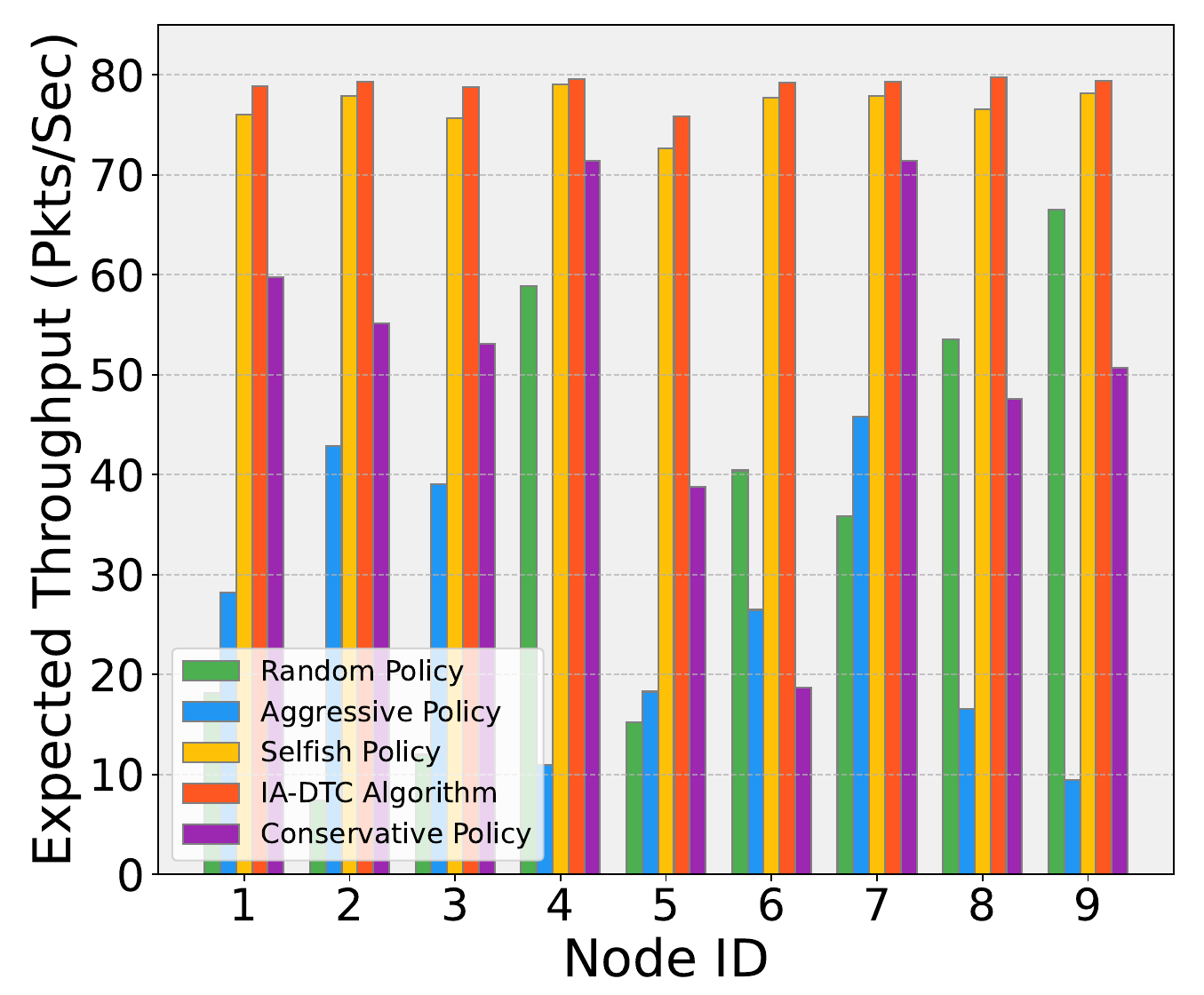}
        \caption{$R_n$ vs. Node ID by IA-DTC}
        \label{fig6}
    \end{minipage}
    \begin{minipage}{.33\textwidth} 
        \includegraphics[width=5.5cm]{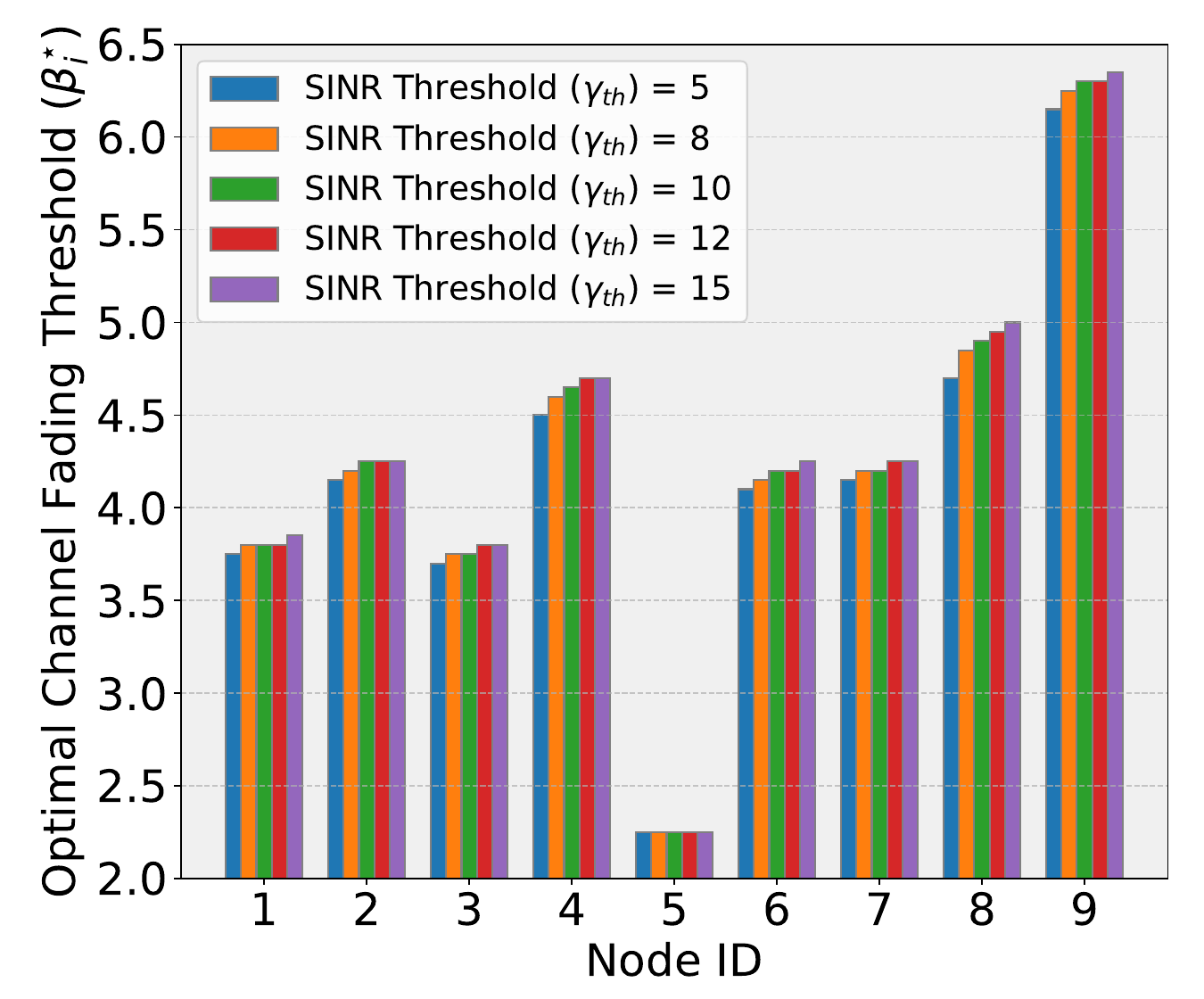}
        \caption{$\beta_i^{\star}$ vs. Node ID by IA-DTC}
        \label{fig7}
    \end{minipage}
    \vspace{-.4cm}
    \label{fig:example}
    \label{IA-DTC1}
\end{figure*}

\textbf{IA-TC Algorithm Performance.} 
The efficacy of the IA-TC algorithm is shown in Figures~\ref{fig2} and \ref{fig3} for three different interferer UAV heights.
From Fig. \ref{fig2}, we observe how the source node finds the optimal $\beta_n$ as the IA-TC algorithm iterates. At the higher altitudes of the interferer UAV (60 m), the
interference impact of the A2A link is larger compared with lower altitudes. Thus, as the altitude of the inteferer UAV increases, the source node chooses larger $\beta_n$ values, thereby enqueuing more packets instead of transmitting to the main UAV. In Fig. \ref{fig2}, initially the source node increases $\beta_n$ since $\boldsymbol{\beta}_m$ are small, but, as the IA-TC algorithm increases $\boldsymbol{\beta}_m$, the source node gradually decreases $\beta_n$, which means that the source node would try transmitting more often due to lower interference. In Fig. \ref{fig3}, we report the throughput performance of the source node as the IA-TC algorithm iterates, which results in a higher $R_n$. Furthermore, $R_n$ increases as the altitude of the interferer UAV decreases, as expected. In the lower altitudes of the interferer UAV (10 m), we achieve the largest $R_n$, since the interferer UAV is located near ground level, its impact is not as significant as the higher altitudes. 







\textbf{IA-DTC Algorithm Performance.} Fig. \ref{fig4} shows that how the optimal channel fading threshold of the source node ($\beta_n^{\star}$) changes as a function of the number of nodes and for different $\gamma_{th}$ values. From the results, we observe that as the number of nodes increases, the source node increases the level of $\beta_n^{\star}$ due to the stronger impact of the interferer nodes. As a result, by increasing the number of nodes and $\gamma_{th}$, the source node acts more conservatively (i.e., fewer transmission attempts) and increases its $\beta_n^{\star}$.

In Fig. \ref{fig5}, we present a comparative analysis of the IA-DTC algorithm output, denoted as $\boldsymbol{\beta}^{\star}$, compared to four baseline transmission policies: 
\textbf{(i) Random policy:} Different nodes select their transmission threshold randomly between zero and the upper bound. \textbf{(ii) Aggressive policy:} Different nodes select their transmission threshold such that the probability of packet drop from the queue (due to overflow or time-threshold) is minimized. In this case, nodes attempt to transmit despite poor channel conditions.  \textbf{(iii) Selfish policy:} Different nodes do not cooperate with each other, and treat other nodes as interference.  \textbf{(iv) Conservative policy:} Different nodes select their transmission threshold to be close to the upper bound values, and thus they aim to minimize outage probability, while increasing the likelihood of packet loss from queues.  
From the results in Fig.~\ref{fig6}, we observe that the IA-DTC algorithm achieves  the highest $R_n$ compared with all other baselines. 
Furthermore, Fig. \ref{fig7} demonstrates the behavior of the IA-DTC algorithm as $\gamma_{th}$ changes. Clearly, as $\gamma_{th}$ increases, nodes increase their optimal channel fading thresholds ($\beta_i^{\star}$) since they require a better channel condition to send their packets, which means that they prefer to enqueue their packets rather than sending them to the UAVs. 
\vspace{-.2cm}
\section{Conclusion} \label{conclusion}
In this paper, we investigated the problem of distributed transmission control for UAVs operating in unlicensed spectrum bands. We developed an analytical interference-aware queuing analysis framework that \emph{jointly} considers three types of packet losses including packet drop due to exceeding the time threshold ($P_n^{dly}(\beta_n)$), buffer overflow ($P_n^{ov}(\beta_n)$), and low SINR ($P_n^{out}(\boldsymbol{\beta}))$.
 Using this analysis, we were able to calculate the throughput performance $R_n$ according to the probability of overall loss. In the transmission policy section, we proposed two algorithms IA-TC and IA-DTC to control $\beta$ for each node to improve $R_n$. We numerically investigated the performance of our algorithms, and confirmed that it achieves the optimal solution. As a future direction, we aim to extend our analysis to model-free analysis and optimization solutions. 

\section*{Acknowledgment}
The material is based upon work supported by NASA under award No(s) 80NSSC20M0261, and NSF grants 1948511, 1955561, and 2212565. Any opinions, findings, and conclusions or recommendations expressed in this material are those of the author(s) and do not necessarily reflect the views of NASA and NSF.

\vspace{-.3cm}
\bibliographystyle{IEEEtran}
\bibliography{ref}

\end{document}